\documentclass[11pt]{article}
\usepackage{amstext,amsmath,amssymb,amsfonts,bbm,amsthm}
\usepackage{graphicx}
\usepackage{authblk}
\usepackage{cite}
\usepackage{epstopdf}

\usepackage[letterpaper, total={6in,8in}]{geometry}

\begin{document}
\title{Thermodynamics of inequalities:\\ from precariousness to economic stratification}
\author{Matteo Smerlak}
\affil{Perimeter Institute for Theoretical Physics\\ 31 Caroline Street North, N2L 2Y5 Waterloo ON, Canada}

\maketitle

\begin{abstract}
Growing economic inequalities are observed in several countries throughout the world. Following Pareto, the power-law structure of these inequalities has been the subject of much theoretical and empirical work. But their \textit{nonequilibrium dynamics}, e.g. after a policy change, remains incompletely understood. Here we introduce a thermodynamical theory of inequalities based on the analogy between economic stratification and statistical entropy. Within this framework we identify the combination of \textit{upward mobility} with \textit{precariousness} as a fundamental driver of inequality. We formalize this statement by a ``second-law" inequality displaying upward mobility and precariousness as thermodynamic conjugate variables. We estimate the time scale for the ``relaxation" of the wealth distribution after a sudden change of the after-tax return on capital. Our method can be generalized to gain insight into the dynamics of inequalities in any Markovian model of socioeconomic interactions.  
\end{abstract}

\section{Introduction}

All known human societies\footnote{Dating back to paleolithic hunter-gatherers \cite{Pringle:2014ke}.} have displayed some level of economic inequality \cite{Piketty:2014cw}. Yet this global imbalance is reaching alarming levels in the contemporary world: as of 2013, the 400 richest Americans have more wealth than the bottom half of all Americans combined. Indeed recent comprehensive research \cite{Piketty:2014uy} has showed that, while they have not reached the highs of the pre-1929 period, wealth inequalities in developed countries have steadily increased in the past decades. Understanding the origins and implications of these inequalities is an outstanding problem for economics, but also for society as a whole. 

On the theory side, a well-established approach to this problem---pursued independently by economists \cite{Champernowne:1953eq}, mathematicians \cite{Mandelbrot:1960eh,Kesten:1973bw}, sociologists \cite{Angle:1986dz} and physicists \cite{Ispolatov:1998kv,Bouchaud:2000ei,Dragulescu:2000bg}---consists in studying the equilibrium wealth distribution in stochastic models of individual (or household) income. Under general assumptions, one shows that additive income lead to exponential distributions, while multiplicative capital returns yield Pareto-like power law distributions \cite{Nirei:2009uo,Anonymous:2005wu,Yakovenko:2009jt,Anonymous:2011hu,Richmond:2013tu}. These results are consistent with empirical data, both contemporary \cite{Dragulescu:2001ds} and historical \cite{AbulMagd:2002ii}, which reveal a two-class structure with an exponential range at low wealth (where investment is negligible) and a power-law tail at high capital (where income is dominated by investment returns). Econophysicists \cite{Dragulescu:2000bg} have pointed the striking similarity between this pattern and the Boltzmann-Gibbs distribution of statistical mechanics. Indeed both have the same ``entropic" structure: there are many more ways to distribute a conserved quantity (be it wealth or energy) unequally than equally.

One much discussed consequence of such marked economic inequalities is the emergence of a super-elite class, the so-called ``top $1\%$" \cite{Atkinson:2009cm}, with disproportionate social, economical and political influence. But they also have more global effects, one of which is increased \textit{stratification} \cite{Grusky:2001tx}---the growth of the number of economically distinct ``classes" in society. Indeed, as we will see below, ``maximum entropy" wealth distributions are precisely those with the greatest stratification under global constraints on the mean wealth. This intriguing analogy between entropy and stratification points to a connection between the dynamics of inequalities and dissipation in thermal systems, extending beyond the limits of equilibrium statistical mechanics (to which it has been restricted so far). 

In this paper we introduce a general framework, inspired from stochastic thermodynamics \cite{Seifert2012}, to account for the dynamical origin of social inequalities. At its foundation is a general property of Markov processes known as the \textit{fluctuation theorem}\footnote{Originally discovered in the context of non-equilibrium statistical mechanics, this result has been successfully applied to models of evolutionary dynamics \cite{Mustonen:2010ig} and of biopoiesis \cite{England:2013ed}. More non-physics applications will likely come in the near future.}  (Appendix \ref{FT}). As we shall see, the great strength of this theorem lies in its \textit{explanatory power}: given an entropy-increasing stochastic process, the fluctuation theorem elucidates the \textit{mechanism} driving entropy production. In the context of social inequalities, where entropy quantifies inequality, we find that, over and above the multiplicative effect of capital return, \textit{precarious social mobility} acts as a universal inequality-generating mechanism.

\section{Results}\label{capital}

\subsection{Stratification}

We begin by formalizing our notion of stratification. Let $w\in[w_{min},w_{max}]$ denote the wealth of an individual (or household) in the economy. The \textit{wealth distribution} $p_t(w)$ is the probability density function (PDF) at time $t$ for the wealth variable $w$, i.e. $p_t(w)dw$ gives the probability of finding an agent with wealth at time $t$ between $w$ and $w+dw$, or the fraction of population whose wealth is between $w$ and $w+dw$ at that time. 

Given $\delta w$ a reference wealth unit, we call economic \textit{stratum} a segment of the population with wealth in the range $[w_i,w_{i+1}]$ where $w_i=w_{min}+b^i\delta w$ for some conventional number $b>1$. For instance, we could take $\delta w=\$1$ and $b=10^3$, in which case the words ``millionaire" and ``billionaire" would correspond to the adjacent strata $i=3$ and $i=4$.

\begin{figure}[t]
\begin{center}
\includegraphics[scale=1]{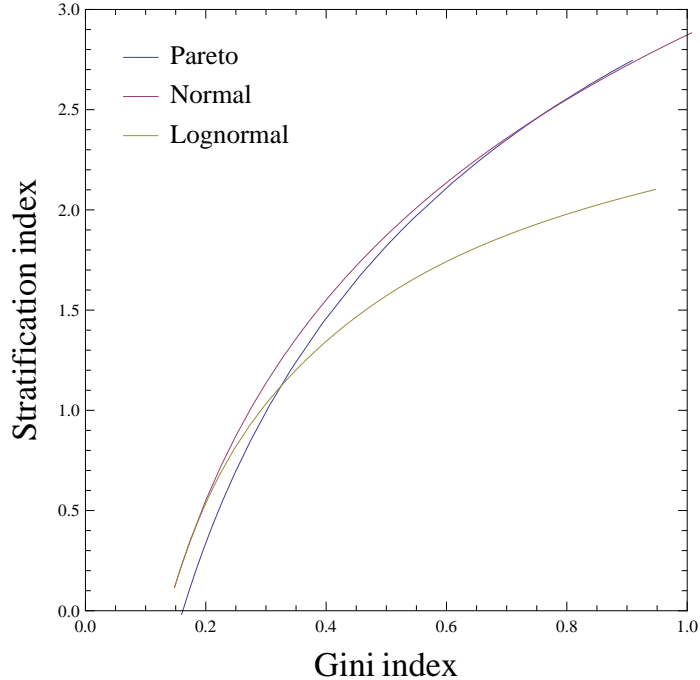}
\end{center}
\caption{Comparison of the Gini and stratification indices for three familiar distributions: Pareto distributions with threshold $w_{min}$ and tail index $\alpha$, normal distributions with mean $m$ and standard deviation $s$, and lognormal distributions with local parameter $\mu$ and scale parameter $\sigma$. Here we fix $w_{min}=m=\mu=1$ and $\delta w=1$ and vary $\alpha$, $s$ and $\sigma$ respectively.}
\label{gini}
\end{figure}

Next we define the \textit{stratification} $S_t$ of the population at time $t$ by
\begin{equation}
  S_t\equiv -\int_{w_{min}}^{w_{min}} p_t(w)\log\,p_t(w)\,dw-\log\delta w,
  \end{equation}
where $\log$ is the base $b$ logarithm. (Mathematically, $S_t$ is the ``differential entropy" of the wealth distribution $p_t(w)$.) Stratification is maximized by the uniform distribution on the interval $[w_{min},w_{max}]$, in which case it simply measures the \textit{number of strata} in the population. This feature is to be contrasted with the Gini index commonly used in the social sciences to measure economic inequalities:
\begin{equation}
G_t\equiv 1-\frac{1}{\langle w\rangle_t}\int_{w_{min}}^{w_{min}}\left[1-\int_{w_{min}}^{w}p_t(w')\,dw'\right]^2\,dw
\end{equation}
where $\langle w\rangle_t$ is the mean of the distribution $p_t(w)$. Indeed, $G_t$ is maximized not by uniform wealth distributions, but by the (highly unrealistic) ``state of extreme inequality" in which one agent has all wealth, and all $N-1$ other agents have nothing: $p(w)=(1-N^{-1})\delta (w-w_{min})+N^{-1}\delta (w-w_{max})$. The fact that the Gini index is maximized by such a singular distribution makes it rather unnatural in the context of large populations with smooth, unimodal distributions. This being said, in many cases of interest the Gini index turns out to be an increasing function of stratification, as illustrated in Fig. \ref{gini}.

It is remarkable that both the Boltzmann (exponential) and Pareto (power-law) distributions,\footnote{The Boltzmann distribution at ``inverse temperature" $\beta$ is $p_B(w)=\beta e^{-\beta w}$, with stratification $1-\log(\beta \delta w)$. The Pareto distribution within minimum wealth $w_{min}$ and Pareto index $\alpha$ is $p_P(w)=\alpha w_{min}^\alpha/w^{\alpha+1}$, with stratification $1+1/\alpha+\log(w_{min}/\alpha \delta w)$.} which have been argued to describe the empirical wealth distributions in the lower and higher quantiles respectively, arise as maximum stratification distributions. Indeed, the former corresponds to the maximum of $S$ under the constraint $\langle w\rangle=1/\beta$, while the latter corresponds to the maximum of $S$ under the constraint $\langle \log (w/w_{min})\rangle=1/\alpha$. (One can check that $S$ is a monotonically decreasing function of $\beta$ and $\alpha$ respectively.) In other words, the lower (resp. higher) quantiles of society appear to be maximally stratified given a fixed mean additive (resp. multiplicative) wealth: using the language of statistical mechanics, we could say that the ``poor" and ``rich" segments of society are close to \textit{statistical equilibrium}\footnote{This notion of equilibrium, which involves no other variable than wealth, should not be confused with other notions of economic equilibrium, such as Nash equilibrium, supply-demand equilibrium or Pareto optimality.} given their constraints. This observation begs a question: how is this social equilibrium reached?

\subsection{A toy model}\label{toy}

\begin{figure}[t!]
\begin{center}
\includegraphics[scale=.8]{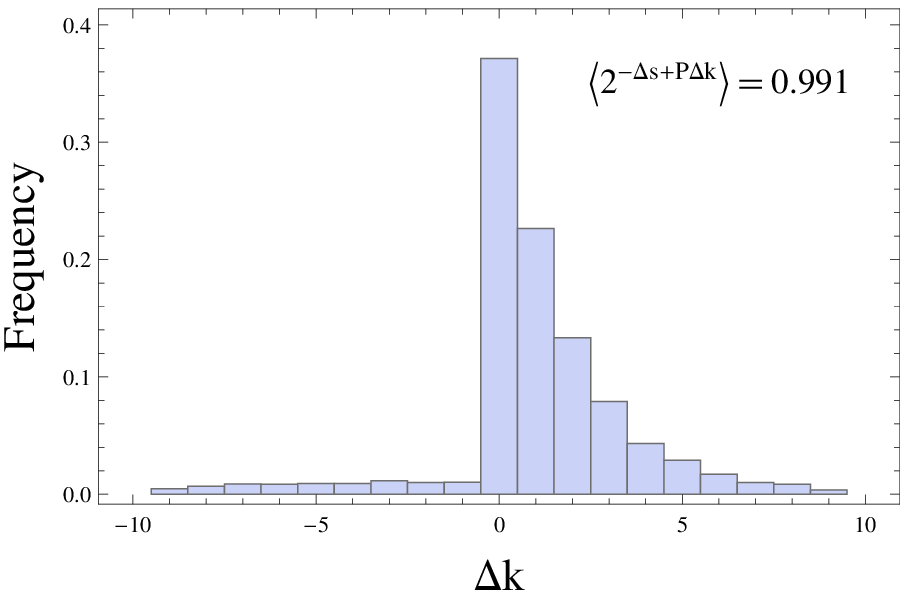}\hfill
\includegraphics[scale=.75]{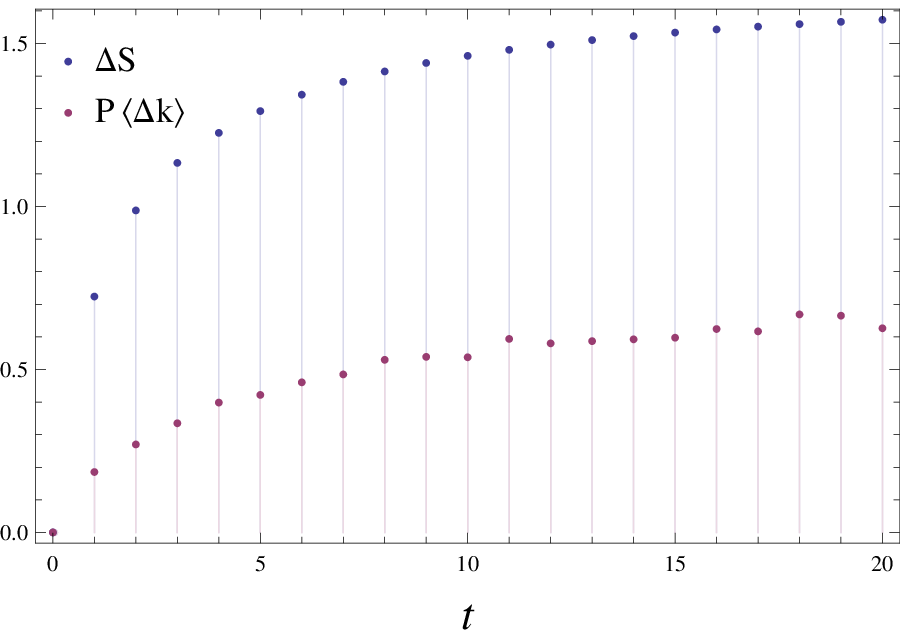}

\bigskip

\includegraphics[scale=.8]{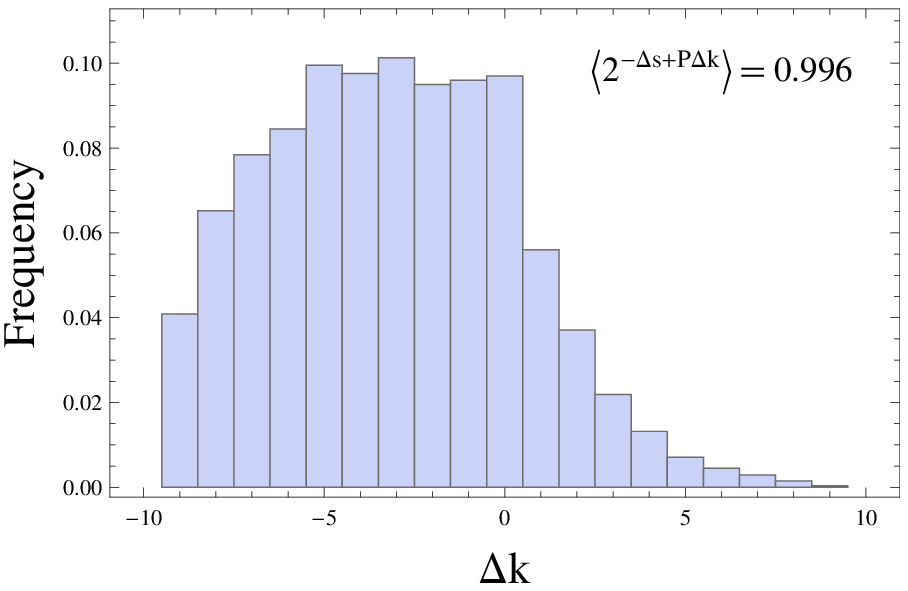}\hfill
\includegraphics[scale=.75]{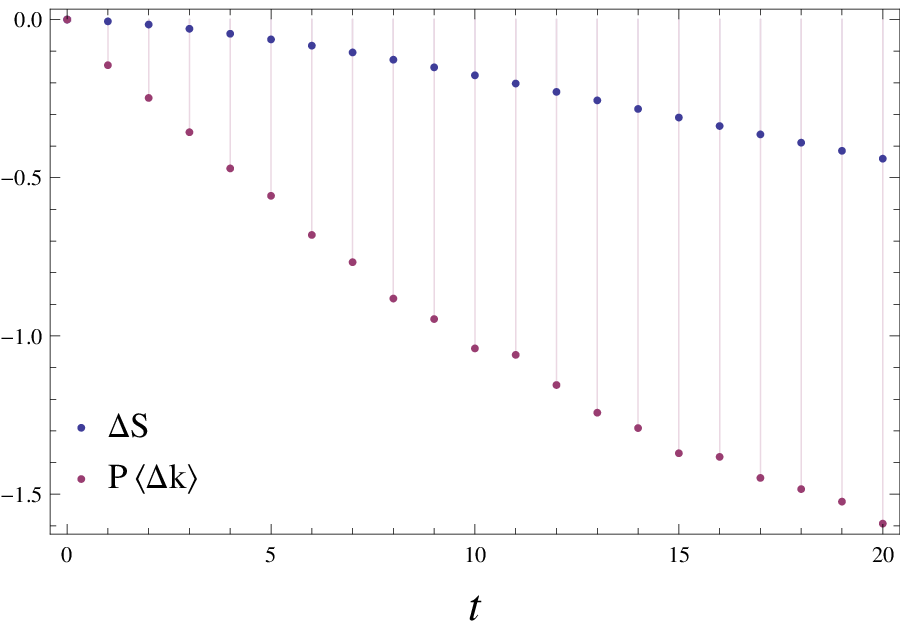}
\end{center}
\caption{Numerical verification of the fluctuation theorem in a toy society with $K=10$ classes, for two different initial distributions $p_0(k)$. In the top row, almost all individuals start in the lowest class ($p_0(k)=91\%$ if $k=1$ and $p_0(k)=1\%$ else). In the bottom row, the initial population is uniformly distribution over the $K$ classes ($p_0(k)=1/K$). Here $\pi_+=30\%$, $\pi_-=50\%$, $b=2$, and the statistics were computed over $10^4$ realizations of the process. The histograms show the distribution of class variations $\Delta k=k_T-k_0$ in the numerical experiment.}
\label{fig2}
\end{figure}

To begin investigating this question, consider the following toy model of society. Assume a finite set of ``classes" $k=1,2,\dots,K$, and suppose that at each time step $t$ there is a probability $\pi_+$ (resp. $\pi_-$) for each individual to move up (resp. down) one class. (Here a ``class" could be an economic stratum in the sense above, or indeed any other form of social ladder: political power, fame, etc.) We call the log-ratio $P=\log(\pi_-/\pi_+)$, which measures of the tendency to go down the social ladder, the \textit{precariousness} parameter. 

The history of an individual in this toy society consists of a Markov chain of classes $k_0$, $k_1$, $k_2$... At each time $t$, the probability to find an individual in class $k$ is the class distribution $p_t(k)$. To this distribution we associate the surprisal $s_t(k)\equiv -\log p_t(k)$; its expectation value is the stratification $S_t\equiv -\sum_k p_t(k)\log p_t(k)$.

Now, the fluctuation theorem for Markov chains (Appendix \ref{chain}) states that, after any given number of time steps $T$, the difference $\Delta s-P\Delta k\equiv s_T(k_T)-s_0(k_0)-P(k_T-k_0)$ is a random variable with the following properties:
\begin{enumerate}
\item
The probability distribution of $\Delta s-P\Delta k$ is such that the expected value
\begin{equation}\label{IFR}
\langle b^{-\Delta s+P\Delta k}\rangle=1.
\end{equation}
This identity implies that $\Delta s-P\Delta k$ is exponentially unlikely to be negative, in the sense that for any positive number $r$,
\begin{equation}\label{negprob}
\textrm{Prob}[\Delta s-P\Delta k\leq -r]\leq b^{-r}.
\end{equation}
\item
As a consequence of this identity, the variation of the stratification $\Delta S=S_T-S_0=\langle \Delta s\rangle$ during the process is constrained by the \textit{second law} inequality
\begin{equation}\label{2ndlaw}
  \Delta S\geq P\langle\Delta k\rangle.
\end{equation}
In other words, in this toy society, 
\begin{equation}
  (\textrm{stratification increase}) \geq (\textrm{precariousness})\times (\textrm{upward mobility}).
\end{equation}

\end{enumerate}
Suppose for instance that all individuals started off in the lowest class $k=1$, so that $S_0=0$, and that $\pi_-<\pi_+$, so that downward social evolution is more likely than upward social evolution. Then the second law indicates that, as the mean class level $\langle k\rangle$ grows, so does the stratification, at a rate greater than $P$ per class level. We illustrate the results \eqref{negprob} and \eqref{2ndlaw} in Fig. \ref{fig2} for two different initial class distributions $p_0(k)$.




\subsection{Stochastic wealth model: ``second law" inequality}\label{stochmodel1}

Armed with this basic intuition, let us now depart from the simplistic notion of ``classes" and get back to continuous wealth distributions. Denote $w_t$ be the detrended\footnote{The detrended wealth is the absolute wealth times $e^{-gt}$ where $g$ is the economic growth rate.} wealth of a household at time $t$. We assume a stochastic dynamics of the form
\begin{equation}\label{model}
  dw_t=l dt+w_t\cdot(\rho dt+\sigma dB_t).
\end{equation}
The first term describe ``additive" income (labor), while the second term represents ``multiplicative" income (capital returns). In the notations of \cite{Piketty:ua}, the mean return rate $\rho$ is given by $\rho=\bar{r}-g-c$, where $\bar{r}$, $g$ and $c$ represent the after-tax return, growth and consumption rates respectively. We assume that the return shocks $dB_t$ form a standard Brownian motion (Wiener process) and we use the Ito convention for stochastic differentials \cite{Gardiner:2004tb}.\footnote{Note that the special case $l=0$ reduces to the geometric Brownian motion widely used in quantitative finance. In this context, our results can be interpreted as putting a lower bound on the uncertainty on an asset price in terms of its drift and volatility.} A dictionary between the terms used in this section and more standard physics terminology is provided in Appendix \ref{dictionary}.

Stochastic wealth models such as \eqref{model} have been considered by many authors, see \cite{Gabaix:2009hc} and references therein. In particular, stochastic equations of the form \eqref{model} arise in the ``random-agent" approximation of certain agent-based models \cite{Boghosian:2014ed}. In this setting one shows \cite{Bouchaud:2000ei} that the stationary distribution---an inverse Gamma distribution---has a Pareto tail with exponent $\alpha=1-2\rho/\sigma^2$, which decreases when $\bar{r}-g$ increases \cite{Piketty:ua}. (When $\rho\geq\sigma^2/2$, the model does not have an equilibrium distribution.)

\begin{figure}\begin{center}
\includegraphics[scale=.8]{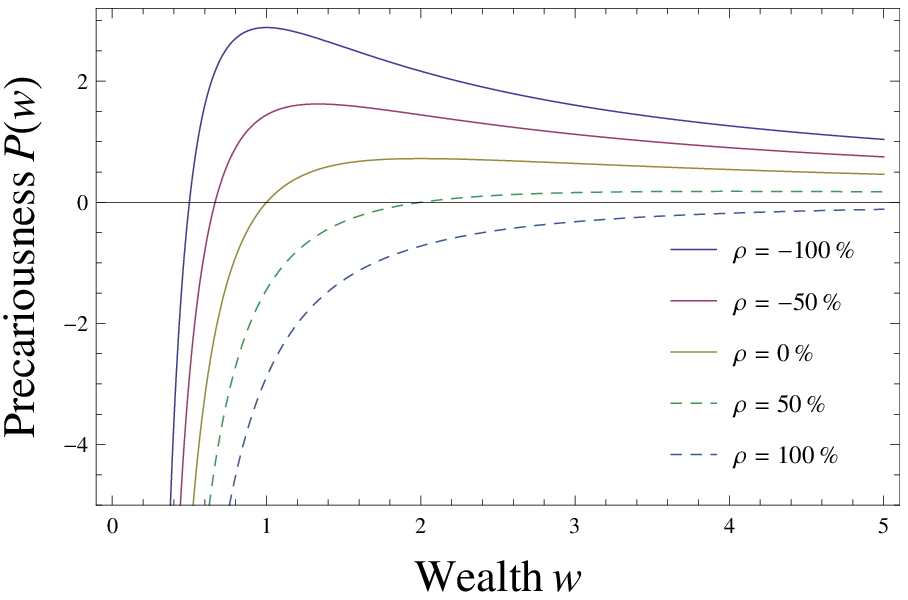}\hfill
\includegraphics[scale=.8]{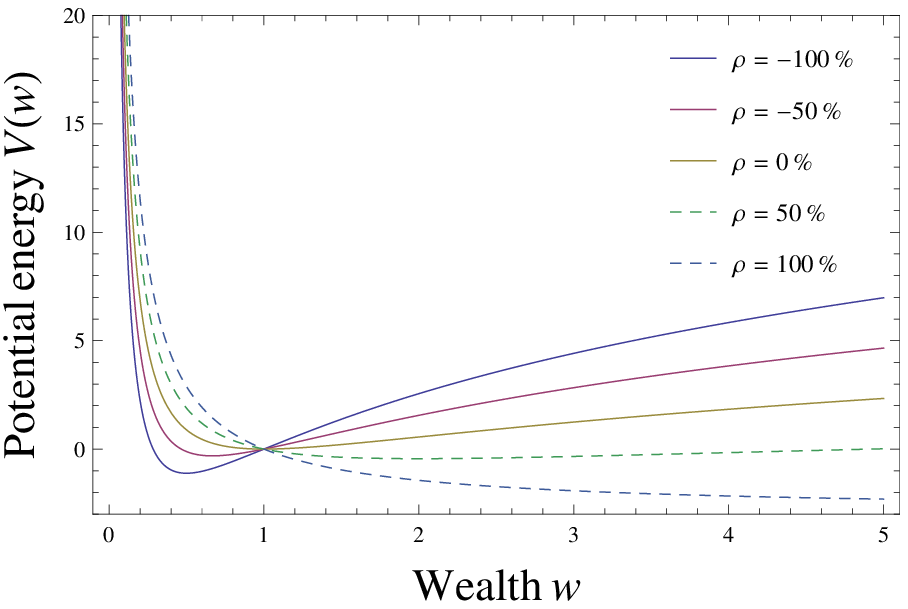}
\end{center}
\caption{The precariousness $P(w)$ (left) and potential $V(w)$ (right) functions for various values of $\rho=\bar{r}-g-c$ at fixed volatility $\sigma$. Here the wealth $w$ is expressed in units of labor income per investment period and we take $\sigma^2=10\%$ per period. The notion of precariousness remains meaningful even in the absence of an equilibrium distribution ($\rho\geq\sigma^2/2$, dashed lines).}
\label{fig3}
\end{figure}

\begin{figure}\begin{center}
\includegraphics[scale=.8]{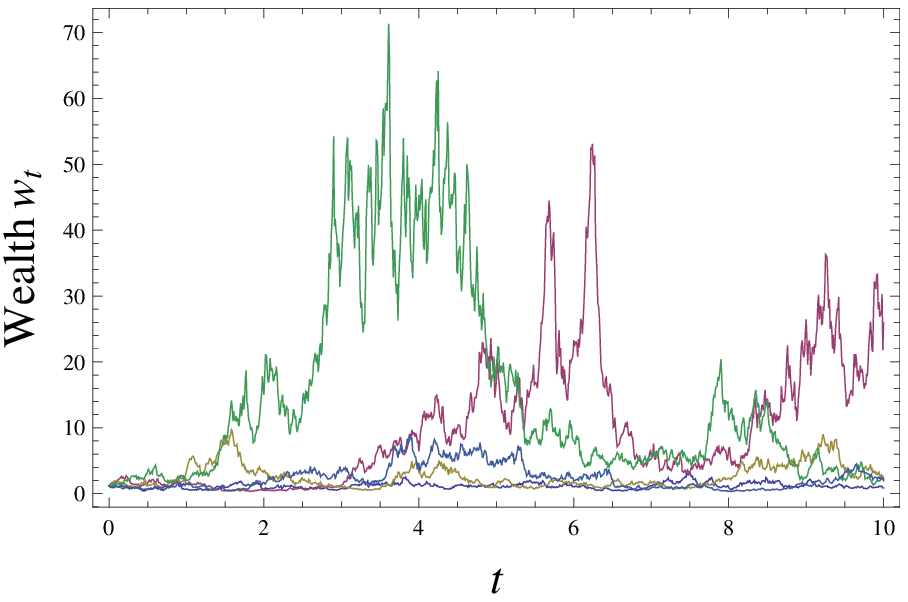}\hfill
\includegraphics[scale=.8]{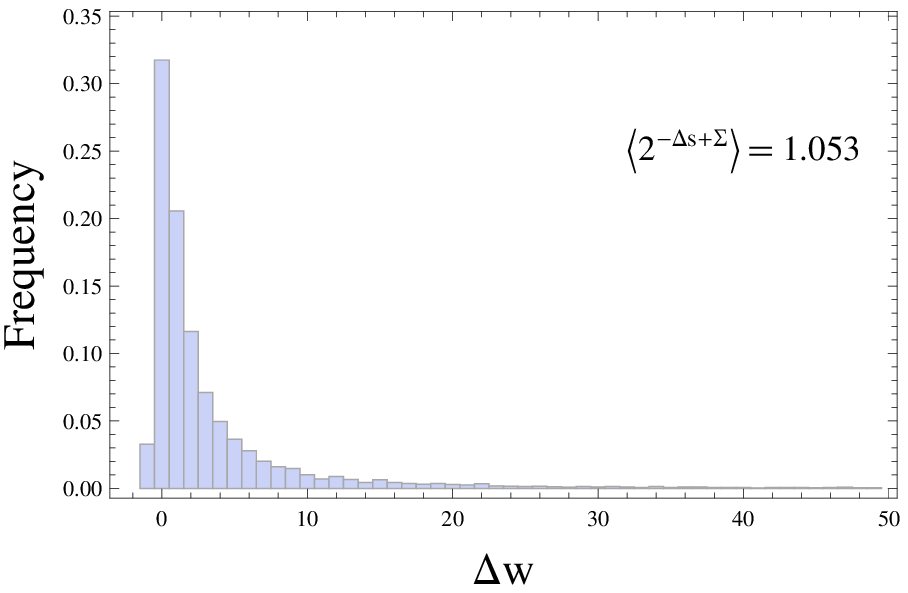}
\end{center}
\caption{Left: five sample paths $w_t$ illustrating the main qualitative feature of the stochastic wealth model \eqref{model}: most paths converge to $w\simeq w_*$ and remain there forever, but some paths make wild excursions at large wealth (the ``top $1\%$" tail of the Pareto distribution). Right: histogram of net gains $\Delta w$ after $T=100$ and verification of the integral fluctuation relation over $10^4$ paths ($\Sigma$ denotes the stochastic integral \eqref{stochint}). In both plots the initial wealth is normally distributed about $w=1$ (with standard deviation $.1$) and $\rho=-10\%$, $l=1$, $\sigma^2=100\%$ per period.}
\label{fig4}
\end{figure}

Here we are interested in the \textit{non-equilibrium} dynamics of stratification. As in the discrete case, this problem can be investigated using the fluctuation theorem for diffusion processes \cite{Lebowitz1999,Maes:2003vh,Seifert2005a}, see Appendix \ref{diff}. This gives the exact analogue of relations \eqref{IFR} and \eqref{negprob}, with the entropy ``source" term $P\Delta k$ replaced by the Stratonovich stochastic integral
\begin{equation}\label{stochint}
\int_0^T P(w_t)\circ dw_t=\int_0^T P(w_t)\cdot dw_t-\frac{1}{2}\int_0^T P'(w_t)dt.
\end{equation}
Here the notations $\cdot\, dw_t$ and $\circ\, dw_t$ refer to the Ito and Stratonovich conventions for stochastic integrals (see e.g. \cite{Gardiner:2004tb}), and we defined the \textit{precariousness function} (plotted in Fig. \ref{fig3})
\begin{equation}\label{defP}
  P(w)=\frac{2}{\ln b}\left(\frac{\sigma^2-\rho}{\sigma^2 w}-\frac{l}{\sigma^2 w^2}\right).
\end{equation}
In particular, the second law inequality now reads\footnote{In this case the stratification rate can be computed exactly, as $$\frac{dS_t}{dt}=\int P(w)j_t(w)dw+\frac{2}{\ln b}\int \frac{j_t(w)^2}{\sigma^2 w^2p_t(w)}dw.$$ In thermodynamic language, the first term is the (reversible) ``entropy flux" and the second term the (irreversible) ``entropy production".}
\begin{equation}\label{2ndlawdiffusion}
  \frac{dS_t}{dt}\geq \int P(w)\,j_t(w)\, dw
\end{equation}
where $j_t(w)$ is the ``social mobility flux", i.e. the expected enrichment rate of a household with wealth $w$. The social mobility flux is given by 
\begin{equation}\label{flux}
 j_t(w)=(l+\rho w)p_t(w)-\frac{\sigma^2}{2}\partial_w(w^2p_t(w))
\end{equation}
and satisfies the continuity (Fokker-Planck, forward Kolmogorov) equation 
\begin{equation}
  \partial_t p_t(w)+\partial_w j_t(w)=0.
\end{equation}
The inequality \eqref{2ndlawdiffusion} becomes an identity in the limit of small flux $j_t(w)$, i.e. close to statistical equilibrium. Fig. \ref{fig4} shows a numerical verification of the fluctuation theorem in this case. 

This result sheds an interesting new light on the relationship between economic conditions and social inequalities. First, when $\rho\leq\sigma^2$, the precariousness function $P(w)$ changes sign at the threshold wealth 
\begin{equation}
w_*=\frac{l}{\sigma^2-\rho}.
\end{equation}
This threshold has a simple interpretation: enrichment $j_t(w)>0$ at $w<w_*$ decreases the stratification, while enrichment $j_t(w)>0$ at $w>w_*$ increases it. This result formalizes the intuitive notion that enriching the poor reduces inequalities, while enriching the rich increase inequalities. 

Second, all other things being equal, this critical wealth $w_*$ is an increasing function of $\bar{r}-g$. This means that a larger capital return rate allows growth to have an inequality-alleviating effect on more quantiles of the population. (This is of course assuming that these quantiles actually participate in multiplicative investments---which of course is not true of the poorer strata of society). Note also that $w_*$, like the Pareto exponent $\alpha$, is a decreasing function of the volatility $\sigma$. Thus, like in our toy model, the growth of inequalities can be interpreted as a consequence of \textit{risk} \cite{Anonymous:2011hu}.  

Third, the precariousness function $P(w)$, which we saw controls the dynamics of stratification out of wealth equilibrium, turns out to be directly related to the equilibrium wealth distribution $p_{\textrm{eq}}(w)$. Indeed, the latter, obtained by setting $j_{\textrm{eq}}=0$ in equation \eqref{flux}, is simply given by
\begin{equation}
  p_{\textrm{eq}}(w)\propto b^{-V(w)}
\end{equation}
where
\begin{equation}
  V(w)\equiv\int^w P(w')dw'.
\end{equation}
The potential function $V(w)$ is plotted in Fig. \ref{fig3}. Thus, from this perspective, the relevant ``conserved quantity" in the economy is $V(w)$, and not wealth\footnote{Indeed, in the model \eqref{model} the equilibrium expected wealth is \textit{infinite} whenever $\rho>0$, i.e. when $\bar{r}-g>c$.} $w$ itself (as proposed in \cite{Dragulescu:2000bg} but criticized in \cite{Hayes:2002tg}). Identifying such a conserved quantity (in physics parlance, a ``potential" function) provides useful intuition for the stochastic dynamics \eqref{model}: roughly speaking, each household tries to reach the minimum of $V(w)$ (the zero-precariousness threshold value $w=w_*$), but is constantly driven away from that value by stratification-maximizing ``fluctuations". In this sense, the potential $V(w)$ can be thought of as the mathematical expression of a Smithian ``invisible hand" driving macroeconomic evolution \cite{Cohen:2013gr}. 

%

\subsection{Stochastic wealth model: relaxation time }

\begin{figure}\begin{center}
\includegraphics[scale=.8]{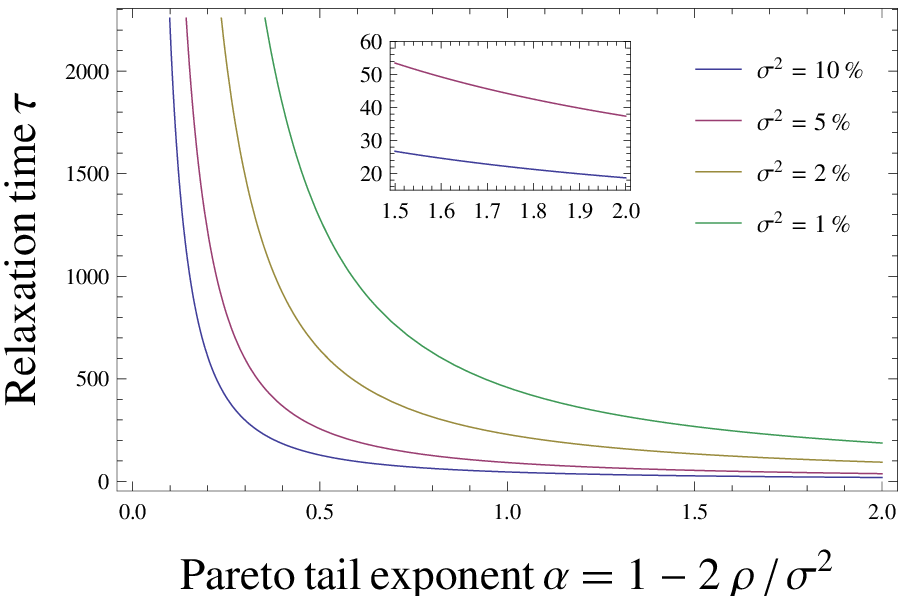}\hfill
\includegraphics[scale=.8]{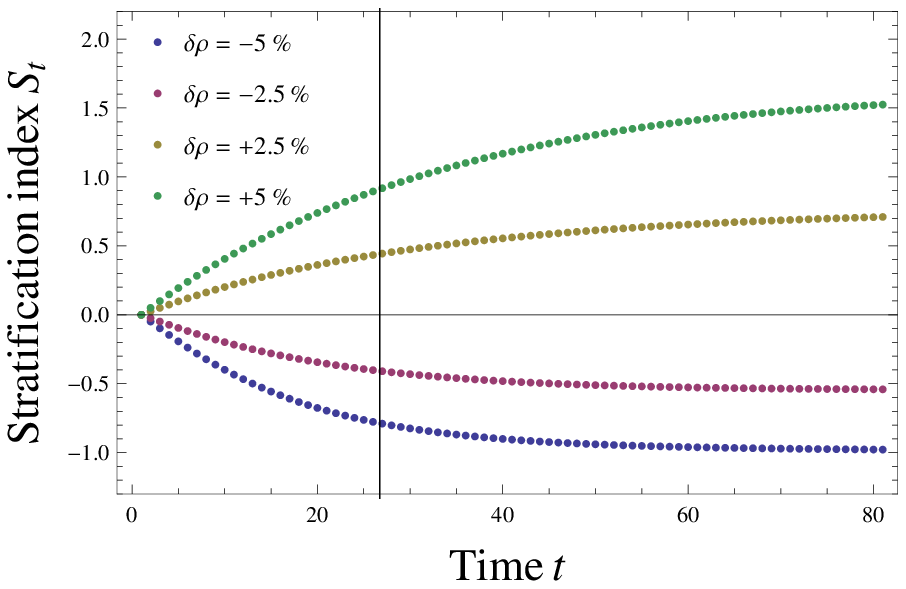}
\end{center}
\caption{Left: the relaxation time \eqref{relaxationtime} as a function of $\alpha=1-2\rho/\sigma^2$. Notice the strong dependence on the volatility $\sigma$ (inset: zoom on the $1.5\leq\alpha\leq 2$ region). Right: Stratification as a function of time, starting from the equilibrium distribution for six different values of $\delta\rho$ (the vertical line represents $t=\tau$). Here $\rho=-2.5\%$, $\sigma^2=10\%$ and $l=1$ per period.}
\label{relax}
\end{figure}

An important question which is readily addressed in this nonequilibrium framework is that of the \textit{relaxation time} of the economy.\footnote{I thank Thomas Piketty for suggesting this problem to me.} Suppose that, starting from the equilibrium wealth distribution for the parameters $(l,\rho,\sigma$) and Pareto tail exponent $\alpha=1-2\rho/\sigma^2$, the detrended effective return rate $\rho$ suddenly changes to the value $\rho'=\rho+\delta\rho$ (e.g. because $\bar{r}-g$ changes according to new fiscal policies): how long will it take for the economy to reach the new statistical equilibrium with Pareto tail exponent $\alpha'=1-2\rho'/\sigma^2$? A straightfoward computation using the formalism above allows us to estimate the relaxation time (to first order in $\delta\rho$) as (Appendix \ref{apprelax})
\begin{equation}\label{relaxationtime}
  \tau\simeq\frac{2\psi_1(\alpha)(1+\alpha)-2}{\sigma^2},
\end{equation}
where $\psi_1$ is the trigamma function. Note that, at this order, the relaxation time $\tau$ is \textit{independent of $\delta\rho$}: larger changes $\delta\rho$ lead to larger changes of the stratification $\delta S$, but these changes always occur on the same time scale $\tau$. The latter is also independent of $l$, implying that increasing labor income does not affect the dynamics of inequalities. Finally, $\tau$ increases with $\bar{r}-g$ and is very sensitive to the volatility $\sigma$, see Fig. \ref{relax}.

\section{Conclusion}

Using the fluctuation theorem for Markov processes as a guide, we have identified \textit{precarious mobility} as a key driver of inequalities away from equilibrium. We have illustrated this idea in simple stochastic wealth models, where we obtained lower bounds on the growth of stratification over time as well as estimates of the corresponding time scales. But the scope of our approach is broader, and can be generalized to other stochastic models of socioeconomic dynamics: in Appendix \ref{exchange} we apply the fluctuation theorem to a model of (biased) trade and find an upper bound on the so-called Theil inequality index. 



Our results are complementary to earlier findings which showed that high interest rates (for given growth rate) generate fat-tailed, Pareto-like equilibrium distributions \cite{Champernowne:1953eq,Mandelbrot:1960eh,Kesten:1973bw,Angle:1986dz,Bouchaud:2000ei,Dragulescu:2000bg,Nirei:2009uo,Anonymous:2005wu,Yakovenko:2009jt,Anonymous:2011hu}. In particular, they throw light on the question: how will the economic system ``respond" to a perturbation? We saw for instance that there exists a threshold wealth $w_*$ beyond which enrichment generates more inequalities. Such knowledge provides a clear guideline regarding the quantiles which should be targeted by redistribution policies ($w<w_*$), and the ones which should not ($w>w_*)$. In the current crisis times, we believe that developing further a ``response theory" of the economic system is a pressing challenge for political economy.

\section*{Acknowledgements}
I was introduced to exchange models of wealth distribution by Tom J. Carter during the Santa Fe Institute CSSS13 summer school. Helpful conversations with Ahmed Youssef are gratefully acknowledged.

\appendix
\section{The fluctuation theorem}\label{FT}

\subsection{General idea}

The (integral) fluctuation theorem is a general property of Markov stochastic processes. It states that there exists a function $\Sigma$ of stochastic paths such that \cite{Seifert2005a} 
\begin{equation}\label{IFRgeneral}
  \langle b^{-\Delta s-\Sigma}\rangle=1,
\end{equation}
where $\Delta s=\log[p_0(X_0)/p_T(X_T)]$ is difference between the initial and final surprisal. Here $X_t$ is the state of the system at time $t$ and $p_t(X)$ is the corresponding probability distribution. The identity \eqref{IFRgeneral} implies that paths such that $\Delta s -\Sigma<0$ are exponentially unlikely, in the sense that if $r$ is a positive number,
\begin{equation}\label{exppos}
  \textrm{Prob}(\Delta s -\Sigma\leq-r)\leq b^{-r}.
\end{equation}
Furthermore, \eqref{IFRgeneral} implies (by convexity of the exponential function) that 
\begin{equation}\label{2ndlawgeneral}
  \langle \Delta s\rangle\geq \langle \Sigma\rangle.
\end{equation}
The left-hand side is nothing but the entropy production $\Delta S=S_T-S_0$, where $S_t$ is the entropy of the distribution $p_t(X_t)$. Thus, the path-dependent function $\Sigma$ can be interpreted as a stochastic entropy source, and \eqref{IFRgeneral} as a refinement of the second law inequality \eqref{2ndlawgeneral}. 

The proof of \eqref{IFRgeneral} is based on the idea of time-reversal. For any stochastic path $X=(X_t)_{t}$ with initial distribution $p_0(X_0)$, consider the time-reversed path $X^\dagger=(X_{T-t})_t$. Denote $\mathbb{P}[X]$ the probability of a path $X$ with initial distribution $p_0(X_0)$, and $\mathbb{P}^\dagger[X]$ the probability of a path with initial  distribution $p_0^\dagger(X_0)=p_T(X_0)$, where $p_T$ is the time-evolution of $p_0$. Next define the path-dependent function $R[X]$ by
\begin{equation}
  R[X]=\frac{\mathbb{P}[X]}{\mathbb{P}^\dagger[X^\dagger]}.
\end{equation}
Then formally
\begin{equation}
  \sum_{X}\mathbb{P}[X]\,b^{-R[X]}=\sum_{X}\mathbb{P^\dagger}[X^\dagger]=1.
\end{equation}
Defining $\Sigma$ by
\begin{equation}
R=\Delta s-\Sigma
\end{equation}
immediately gives \eqref{IFRgeneral}. To gain useful information about entropy production in a given Markov process, it therefore suffices to compute explicitly the log-ratio $R$. 

Equation \eqref{exppos} an immediate consequence of \eqref{IFRgeneral} \cite{jarzynski}:
\begin{eqnarray}
\textrm{Prob}(\Delta s-\Sigma\leq-r)&=&\int_{-\infty}^{-r}\textrm{Prob}(\Delta s-\Sigma=q)\,dq\\
&\leq &\int_{-\infty}^{-r}\textrm{Prob}(\Delta s-\Sigma=q)\, b^{-q-r}\,dq\\
&\leq &2^{-r}\int_{-\infty}^{+\infty}\textrm{Prob}(\Delta s-\Sigma=q)\, b^{-q}\,dq\\
&\leq &b^{-r}.
\end{eqnarray}
We outline below the computation of $R=\Delta s-\Sigma$ for discrete states (sec. \ref{chain}) and diffusions (\ref{diff}) in the stationary case; the nonstationary case can be treated along the exact same lines.

\subsection{Markov chains}\label{chain}

Let us begin by considering a discrete-space, discrete-time Markov chain. Let $\gamma_{ij}$ be the transition probability between states $i$ and $j$, $p_0(i)$  the initial probability distribution, and $p_T(i_T)$ the final probability distribution after $T$ time steps.\footnote{The two are related by the matrix equation $p_T=\Gamma^T p_0$, where $\Gamma$ is the matrix with entries $\gamma_{ji}$} Then the probability of a path $(i_0,i_1,\cdots,i_N)$ is given by 
\begin{equation}
\mathbb{P}(i_0,i_1,\cdots,i_N)=p_0(i_0)\prod_{k=0}^{N-1}\gamma_{i_ki_{k+1}}
\end{equation}
and the probability of the reverse path $(i_N,i_{N-1},\cdots,i_0)$ with initial distribution $p_0^\dagger(i_N)=p_T(i_N)$ is 
\begin{equation}
\mathbb{P}^\dagger(i_N,i_{N-1},\cdots,i_0)=p_T(i_T)\prod_{k=0}^{N-1}\gamma_{i_{k+1}i_{k}}.
\end{equation}
Hence 
\begin{equation}
  R(i_0,i_1,\cdots,i_N)=\log\frac{p_0(i_0)}{p_T(i_T)}+\sum_{k=0}^{N-1}\log\frac{\gamma_{i_ki_{k+1}}}{\gamma_{i_{k+1}i_k}}.
\end{equation}
The first term is the surprisal difference $\Delta s$, and the second term defines the entropy source as
\begin{equation}\label{sigma}
  \Sigma=\sum_{k=0}^{N-1}\log\frac{\gamma_{i_{k+1}i_{k}}}{\gamma_{i_{k}i_{k+1}}}.
\end{equation}
The structure of $\Sigma$ provides a clear-cut explanation for the origin of entropy production: on average, entropy grows when the system makes transitions $i_k\rightarrow i_{k+1}$ which are disfavored with respect to the reverse transitions $i_{k+1}\rightarrow i_{k}$.

This result immediately generalizes to continuous-time Markov chains. In that case, $\gamma_{ij}$ are transition rates rather than probabilities and the probability of a path $(i_0,i_1,\cdots,i_N)$ with transition times $(t_1,\cdots,t_N)$ is given by 
\begin{equation}
P(i_0,i_1,\cdots,i_N;t_1,\cdots,t_N)=p_0(i_0)\prod_{k=0}^{N-1}e^{-\lambda_k(t_{k+1}-t_k)}\gamma_{i_ki_{k+1}}.
\end{equation}
where $\lambda_k=\sum_j \gamma_{i_kj}$. The log-ratio $R$ is unchanged, and $\Sigma$ is still given by \eqref{sigma}.

\subsection{Diffusion processes}\label{diff}

Consider now an Ito diffusion process
\begin{equation}\label{forward}
  dX_t=c(X_t)\,dt+d(X_t)\cdot dB_t
\end{equation}
with initial probability density $p_0(X_0)$ and path measure $d\mathbb{dP}[X]$. Here $dB_t$ is a standard Wiener process. Introduce the ``time-reversed" process $X_t^\dagger=X_{T-t}$, with equation
\begin{equation}
  dX_t^\dagger= -c(X_t^\dagger)\,dt+d(X_t^\dagger)\cdot dB_t.
\end{equation}
and initial probability $p_0^\dagger(X^\dagger_0)=p_T(X^\dagger_0)$, where $p_T$ is the time-evolution of $p_0$.\footnote{The final distribution $p_T$ is the solution of the forward Kolmogorov equation with initial condition $p_0$.} Denote $d\mathbb{P}^\dagger[X^\dagger]$ its path measure. Then, by the Girsanov theorem, the Radon-Nikodym derivative of $d\mathbb{dP}$ with respect to $d\mathbb{dP^\dagger}$ satisfies \cite{Lebowitz1999}
\begin{equation}
\ln\frac{d\mathbb{P}[X]}{d\mathbb{P}^\dagger[X^\dagger]}=\Delta s+2\int_0^T \left(\frac{c(X_t)-d(X_t)d'(X_t)}{d^2(X_t)}\right)\circ dX_t
\end{equation}
where the Stratonovich integral is defined by
\begin{equation}
 \int_0^Tf(X_t)\circ dX_t= \int_0^Tf(X_t)\cdot dX_t-\frac{1}{2}\int_0^T f'(X_t)dt.
\end{equation}
The entropy source function is thus given in this case by 
\begin{equation}
  \Sigma=-\frac{2}{\ln 2}\int_0^T \left(\frac{c(X_t)-d(X_t)d'(X_t)}{d^2(X_t)}\right)\circ dX_t.
\end{equation}

%
%
%
%

\section{Inequalities from biased trade}\label{exchange}

\subsection{Theil index}

The entropic definition of stratification is reminiscent of the Theil inequality index. The two metrics, however, are conceptually different: unlike stratification, the Theil index involves $N$  different agents $n$ with wealth $w_t^{n}$ at time $t$; it is defined by 
\begin{equation}
  T_t=\sum_{n=1}^N \phi^{n}_t\log\phi^{n}_t+\log N
\end{equation}
where $\phi_t^n=w_t^n/\sum_{n=1}^N w_t^{n}$ is the fraction of the total wealth held by agent $n$. Formally, the Theil index is the difference between the maximal and observed Shannon entropy of the distribution of wealth fractions. Remarkably, the fluctuation theorem provides insight also in the dynamics of the Theil index, albeit in a dual way with respect to stratification: as we now show, the fluctuation theorem provides an \textit{upper} bound on the growth of $T_t$.

\subsection{An exchange model}

Suppose that $N$ agents trade their wealth among themselves, following the rule that $n$ transfers to $m$ an amount (proportional to his wealth $w_t^n$) at the rate $J_{nm}$. The exchange rates $J_{nm}$ need not be symmetric; we define the trade \textit{bias} from $n$ to $m$ by $b_{nm}=\log(J_{nm}/J_{mn})$. The wealth fractions are governed by the rate equations \cite{Bouchaud:2000ei}
\begin{equation}
\frac{d\phi^{n}_t}{dt}=\sum_{m\neq n}\left(J_{mn}\phi_t^{m}-J_{nm}\phi_t^{n}\right).
\end{equation}

Importantly, these equations can be interpreted as the master equations of a continuous-time Markov chain $n_t$. In this interpretation, wealth can be thought of as a token that randomly changes hands among the agents; $\phi_t^n$ is then the probability to find the token with agent $n$ at time $t$. Applying the fluctuation theorem to this chain, we obtain the statistical constraint \eqref{IFR}, now with $-B=-\sum_t b_{n_t n_{t+1}}$ as a source term (playing the role of $P\Delta k$ in sec. \ref{toy}). The second law inequality takes the form

\begin{equation}\label{Theillaw}
  \Delta T\leq \langle B\rangle,
\end{equation}
i.e. the increase of the Theil index is smaller than the expected cumulated bias along the chain. In particular, if all exchanges are unbiased, we obtain the intuitive result that inequalities must decrease over time.

To illustrate this result, consider a trading society with just three agents $f_1$, $f_2$ and $u$ ($f$ standing for ``fair" and $u$ for ``unfair"). Assume that $J_{f_1f_2}=J_{f_2f_2}=J_0$, $J_{f_1u}=J_{f_2u}=J_+$ and $J_{uf_1}=J_{uf_2}=J_-$ with $b=\log(J_+/J_-)>0$. Then the second law \eqref{Theillaw} implies that, while the presence of $u$ clearly creates inequalities, the Theil index cannot grow more than  $b$ times the mean number of unreciprocated exchanges from $f_{1,2}$ to $u$. 

\section{Computation of the relaxation time \eqref{relaxationtime}}\label{apprelax}

\begin{figure}\begin{center}
\includegraphics[scale=1]{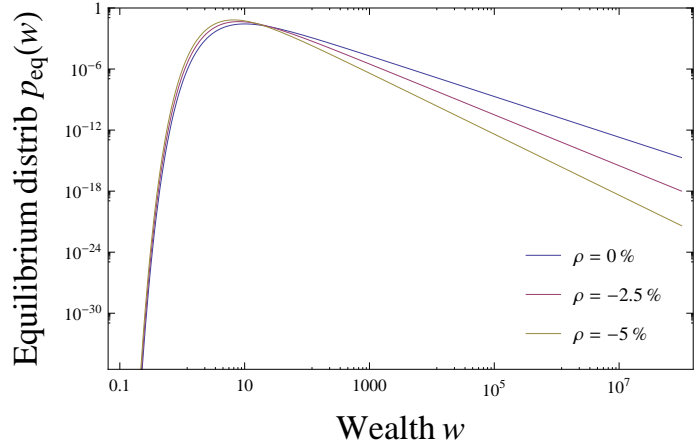}\end{center}
\caption{Equilibrium wealth distribution $p_{\textrm{eq}}(w)\propto b^{-V(w)}$ of the stochastic model \eqref{model} for various values of $\rho=\bar{r}-g-c$, corresponding to different Pareto tail exponent $\alpha=1-2\rho/\sigma^2$. Here $l=1$ and $\sigma^2=10\%$ per period.}
\label{equi}
\end{figure}

The equilibrium distribution $p_{\textrm{eq}}(w)$ for the stochastic model \eqref{model} is the inverse gamma distribution
\begin{equation}\label{eq}
  p_{\textrm{eq}}(w)=\frac{\beta^{\alpha}e^{-\beta/w}
}{\Gamma(\alpha)\,w^{\alpha+1}}\end{equation}
where $\alpha=1-2\rho/\sigma^2$ and $\beta=2l/\sigma^2$ (Fig. \ref{equi}). Its differential entropy is given by
\begin{equation}\label{entropyinversegamma}
  -\int_0^{\infty}p_{\textrm{eq}}(w)\log p_{\textrm{eq}}(w)\,dw=\frac{\alpha+\ln(2l/\sigma^2)+\ln \Gamma(\alpha)-(1+\alpha)\psi(\alpha)}{\ln b},
\end{equation}
where $\Gamma$ and $\psi$ and the gamma and digamma functions. Now, suppose that at time $t=0$, the effective return rate $\rho=\bar{r}-g-c$ changes to a different value $\rho'=\rho+\delta\rho$. We can estimate the time before the the wealth distribution settles to a new equilibrium with tail exponent $\alpha'=1-2\rho'/\sigma^2$ as 
\begin{equation}\label{deftau}
  \tau\simeq\left(\frac{\partial S_{\textrm{eq}}}{\partial\rho}\,\delta\rho\right)\Big/\left(\frac{dS_t}{dt}\right)\Big|_{t=0}.
\end{equation}
From \eqref{entropyinversegamma} we compute
\begin{equation}\label{comp1}
  \frac{\partial S_{\textrm{eq}}}{\partial\rho}=\left(\frac{-2}{\sigma^2}\right)\left(\frac{1-\psi_1(\alpha)(1+\alpha)}{\ln b}\right).
\end{equation}
Next we estimate the denominator of \eqref{deftau} using 
\begin{equation}
  \frac{dS_t}{dt}\Big|_{t=0}\simeq\int_0^{\infty}P(w)\,j_{0}(w)dw.
\end{equation}
where the initial flux $j_0(w)$ is given by
\begin{eqnarray}
  j_0(w)&=&(l+\rho'w)p_{\textrm{eq}}(w)-\frac{\partial}{\partial w}\left(\frac{\sigma^2w^2p_{\textrm{eq}}}{2}\right)\\
  &=&(l+\rho'w)p_{\textrm{eq}}(w)-(l+\rho w)p_{\textrm{eq}}(w)\\
  &=&(\delta\rho\, w) p_{\textrm{eq}}(w).
\end{eqnarray}
This gives
\begin{equation}
  \frac{dS_t}{dt}\Big|_{t=0}\simeq\delta\rho\int_0^{\infty}wP(w)\, p_{\textrm{eq}}(w)dw.
\end{equation}
The integral on the right-hand side can be evaluated explicited using \eqref{defP} and \eqref{eq}, yielding
\begin{equation}\label{comp2}
  \int_0^{\infty}wP(w)\, p_{\textrm{eq}}(w)dw=\frac{1}{\ln b}.
\end{equation}
Plugging \eqref{comp1} and \eqref{comp2} into \eqref{deftau} gives \eqref{relaxationtime}.

\section{Dictionary}\label{dictionary}

Here we provide for the reader's convenience a dictionary relating the concepts used in this paper, notably in sec. \ref{stochmodel1}, to their physics counterparts.  

\medskip

\begin{center}
\begin{tabular}{|c|c|}
  \hline
  Economics & Physics \\
  \hline
  Stratification & Gibbs entropy  \\
  Precariousness & (Driving force)/(temperature)\\
  Mobility flux & Probability current\\
  Potential & Potential energy\\
  Surprisal & Stochastic or local entropy\\
  \hline
\end{tabular}
\end{center}

\bigskip

\bibliographystyle{utcaps}
\providecommand{\href}[2]{#2}\begingroup\raggedright\endgroup

\end{document}